\documentclass[twoside]{article}

\usepackage{PRIMEarxiv}

\usepackage[utf8]{inputenc} 
\usepackage[T1]{fontenc}    
\usepackage{hyperref}       
\usepackage{url}            
\usepackage{booktabs}       
\usepackage{amsfonts}       
\usepackage{nicefrac}       
\usepackage{microtype}      
\usepackage{lipsum}
\usepackage{fancyhdr}       
\usepackage{graphicx}       
\graphicspath{{media/}}     

\title{GENHOP: An Image Generation Method Based on Successive Subspace Learning
\thanks{\textit{\underline{Cite as}}: 
\textbf{Xuejing Lei, Wei Wang and C.-C. Jay Kuo, “GENHOP: an image generation method based on successive subspace learning,” IEEE International Symposium on Circuits \& Systems (ISCAS), Austin, Texas, USA May 28-June 1, 2022.}}
}

\author{
  Xuejing~Lei\\
  Media Communications Lab\\
  University of Southern California\\
  Los Angeles, CA, USA \\
  \texttt{xuejing@usc.edu} \\
  \And
  Wei~Wang \\
  Media Communications Lab\\
  University of Southern California\\
  Los Angeles, CA, USA \\
  \texttt{wang890@usc.edu} \\
  \AND
  C.-C.~Jay~Kuo \\
  Media Communications Lab\\
  University of Southern California\\
  Los Angeles, CA, USA \\
  \texttt{jckuo@usc.edu} \\
}

\begin{document}
\maketitle

\begin{abstract}
Being different from deep-learning-based (DL-based) image generation
methods, a new image generative model built upon successive subspace
learning principle is proposed and named GenHop (an acronym of
Generative PixelHop) in this work.  GenHop consists of three modules: 1)
high-to-low dimension reduction, 2) seed image generation, and 3)
low-to-high dimension expansion.  In the first module, it builds a
sequence of high-to-low dimensional subspaces through a sequence of
whitening processes, each of which contains samples of
joint-spatial-spectral representation.  In the second module, it
generates samples in the lowest dimensional subspace. In the third
module, it finds a proper high-dimensional sample for a seed image by
adding details back via locally linear embedding (LLE) and a sequence of
coloring processes. Experiments show that GenHop can generate visually
pleasant images whose FID scores are comparable or even better than
those of DL-based generative models for MNIST, Fashion-MNIST and CelebA
datasets. 
\end{abstract}

\keywords{Image Generative Modeling \and Image Generation \and Successive Subspace Learning}

\section{Introduction}\label{sec:intro}

Unconditional image generation has received increasing attention
recently due to impressive results offered by deep-learning (DL) based
methods such as generative adversarial networks (GANs), variational
auto-encoders (VAEs), and flow-based methods. Yet, DL-based methods are
blackbox tools.  The end-to-end optimization of networks is a non-convex
optimization problem, which is mathematically intractable. Being
motivated by the design of other generative models that allow
mathematical interpretation, a new image generative model is proposed in
this work.  Our method is developed based on the successive subspace
learning (SSL) principle~\cite{kuo2016understanding, kuo2017cnn,kuo2018data, kuo2019interpretable} and built upon the foundation of the
PixelHop++ architecture~\cite{chen2020pixelhop++}. Thus, it is called
Generative PixelHop (or GenHop in short). Its high-level idea is
sketched below. 

Since high-dimensional input images have complicated statistical
correlations among pixel values, it is difficult to generate images
directly in the pixel domain. To address this problem, GenHop contains
three modules: 1) high-to-low dimension reduction, 2) seed image
generation, and 3) low-to-high dimension expansion. In the first module,
it builds a sequence of high-to-low dimensional subspaces through a
sequence of whitening processes called the channel-wise Saab transform,
where high frequency components are discarded to lower the dimension.
In the second module, the sample distribution in the lowest dimensional
subspace can be analyzed and generated by white Gaussian noise,
which is called seed image generation.  In the third module, GenHop
attempts to find the corresponding source image of a seed image through
dimension expansion and a coloring mechanism. For dimension expansion,
discarded high frequency components are recovered via locally linear
embedding (LLE).  The coloring process is the inverse of the whitening
process, which is achieved by the inverse Saab transform.  Experiments
are conducted on MNIST, Fashion-MNIST and CelebA three datasets to
demonstrate that GenHop can generate visually pleasant images whose FID
scores are comparable with (or even better than) those of DL-based
generative models. 

\section{Review of Related Work}\label{sec:review}

{\bf DL-based Generative Models.} An image generative model learns the distribution of image samples from
a certain domain and then generates new images that follow the learned
distribution. Generally speaking, the design of image generative models
involves analysis and generation two pipelines. The former analyzes
properties of training image samples while the latter generates new
images after the training is completed. Only the generation unit is used
for image generation in inference. So far, the best performing image
generative models are all DL-based.  We may categorize DL-based
generative methods into two categories: adversarial and non-adversarial
models.  For the adversarial category, generative adversarial networks
(GANs)~\cite{goodfellow2014generative} demand that distributions of
training and generated images are indistinguishable. This can be achieved
by training a generator/discriminator pair through end-to-end
optimization of a certain cost function. GANs exhibit good
generalization capability and yield visually impressive images.  For the
non-adversarial category, examples include Variational Auto-Encoders
(VAEs)~\cite{Kingma2014}, flow-based methods~\cite{dinh2014nice,dinh2016density} and GLANN~\cite{hoshen2019non}. VAEs learns an
approximation of the density function with an encoder/decoder structure.
Flow-based methods transform the Gaussian distribution into a complex
distribution by applying a sequence of invertible transformation
functions.  GLANN~\cite{hoshen2019non} maps images to a feature space
obtained by GLO~\cite{bojanowski2018optimizing} and maps the feature
space to the noise space via IMLE~\cite{li2018implicit}. It achieves the
state-of-the-art performance among non-adversarial methods. 

{\bf SSL.} Traditional spectral analysis such as the Fourier transform and the
principle component analysis (PCA) attempts to capture the global
structure but sacrifices local detail (e.g., object boundaries) of
images. In contrast, local detail can be well described in the spatial
domain, yet the pure spatial representation cannot capture the global
information well. To overcome these shortcomings, Kuo {\em et al.}
~\cite{kuo2016understanding, kuo2017cnn, kuo2018data,
kuo2019interpretable} proposed two affine transforms that determine a
sequence of joint spatial-spectral representations of different
spatial/spectral trade-offs to characterize the global structure and
local detail of images at the same time. They are the Saak transform
\cite{kuo2018data} and the Saab transform \cite{kuo2019interpretable}.
As a variant of the Saab transform, the channel-wise (c/w) Saab
transform \cite{chen2020pixelhop++} exploits weak correlations among
spectral channels and applies the Saab transform to each channel
individually to reduce the model size without sacrificing the
performance. The mathematical theory is named successive-subspace-learning (SSL).
PixelHop~\cite{chen2020pixelhop} and
PixelHop++~\cite{chen2020pixelhop++} are two architectures developed to
implement SSL.  PixelHop consists of multi-stage Saab transforms in
cascade. PixelHop++ is an improved version of PixelHop by replacing the Saab transform with the c/w Saab transform. SSL has been successfully applied to many application domains. Examples
include~\cite{chen2021defakehop,zhang2021anomalyhop,kadam2021r,liu2021voxelhop, zhang2020pointhop++,zhang2020pointhop,zhang2020unsupervised, kadam2020unsupervised,manimaran2020visualization, tseng2020interpretable,rouhsedaghat2020facehop, rouhsedaghat2021successive}. It is worthwhile to mention that SSL-based texture synthesis was studied in~\cite{9306451,lei2021tghop}. Here, we examine SSL-based image generation that goes beyond texture and demands a few extensions such as improved fine detail generation, quality enhancement via local detail generation, etc.

\begin{figure*}[tb]
\centering
\includegraphics[width=1\linewidth]{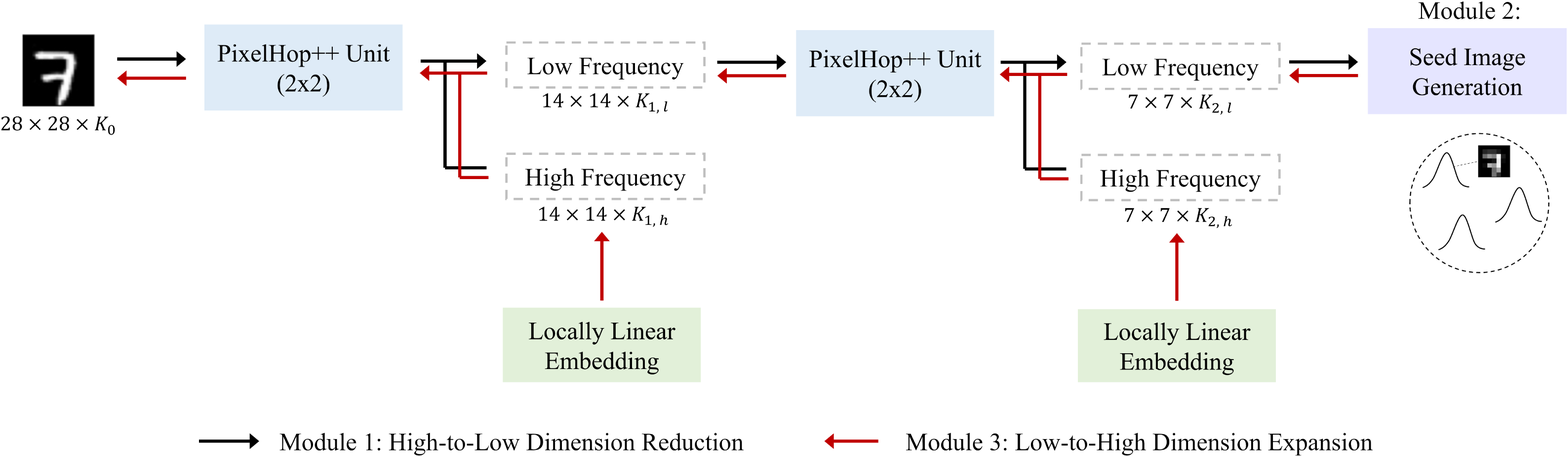}
\caption{An overview of the GenHop method. Four subspaces
$S_1$, $S_2$, $S_3$ and $S_4$ are constructed from source image space $S_0$ with two PixelHop++ units. GenHop contains three modules: 1) High-to-Low Dimension Reduction, 2) Seed Image Generation and 3) Low-to-High Dimension Expansion.}\label{fig:intro}
\end{figure*}

\section{Proposed GenHop Method}\label{sec:genhop}

An overview of the proposed GenHop method is shown in Fig.
\ref{fig:intro}, which contains three modules as elaborated below.

\vspace{-1ex}
\subsection{Module 1: High-to-Low Dimension Reduction}\label{subsec:analysis}

A sequence of high-to-low dimensional subspaces is constructed from the
source image space $S_0$ through PixelHop++ \cite{chen2020pixelhop++} as
shown in the figure. Each PixelHop++ unit behaves like a whitening
operation. It decouples a local neighborhood (i.e., a block) into DC and
AC parts and conducts the principal component analysis (PCA) on the AC
part. This is named the Saab transform. The reason to remove the DC
first is that the ensemble mean of AC part can be well approximated by
zero so that the PCA can be applied without the need to estimate the
ensemble mean. The PCA is essentially a whitening process. It removes
the correlation between AC components among pixels in the same block. 

To give an example, for an input gray-scale image of size 28x28, we
apply the Saab transform to 2x2 non-overlapping blocks, which offers one
DC and three AC channels per block, in the first PixelHop unit.  The
output is a joint-spatial-spectral representation of dimension 14x14x4,
which forms subspace $S_1$. We can partition spectral channels into low-
and high-frequency channels whose numbers are denoted by $K_{1,l}$ and
$K_{1,h}$, respectively. Low-frequency channels have larger energy
representing the main structure of an image while high-frequency
channels has lower energy representing image details. Only low-frequency
channels proceed to the next stage. In other words, high-frequency
channels are discarded to lower the dimension and will be estimated via LLE as discussed in Sec. \ref{subsec:generation}. High-frequency channels with sufficiently small energy will not be estimated, for example, on MNIST dataset. This lead to the sum of $K_{1,l}$ and $K_{1,h}$ being less than 4. The cascade of several PixelHop++ units yields several subspaces.  For images of small spatial resolutions, we adopt two PixelHop++ units as shown in Fig. \ref{fig:intro} to ensure a proper spatial resolution in $S_4$, which has the lowest joint spatial-spectral dimension, to capture the global structure of an image. 

\subsection{Module 2: Seed Image Generation} \label{subsec:seed}

\begin{figure}[bt]
\centering
\includegraphics[width=\linewidth, height=0.8\textheight, keepaspectratio]{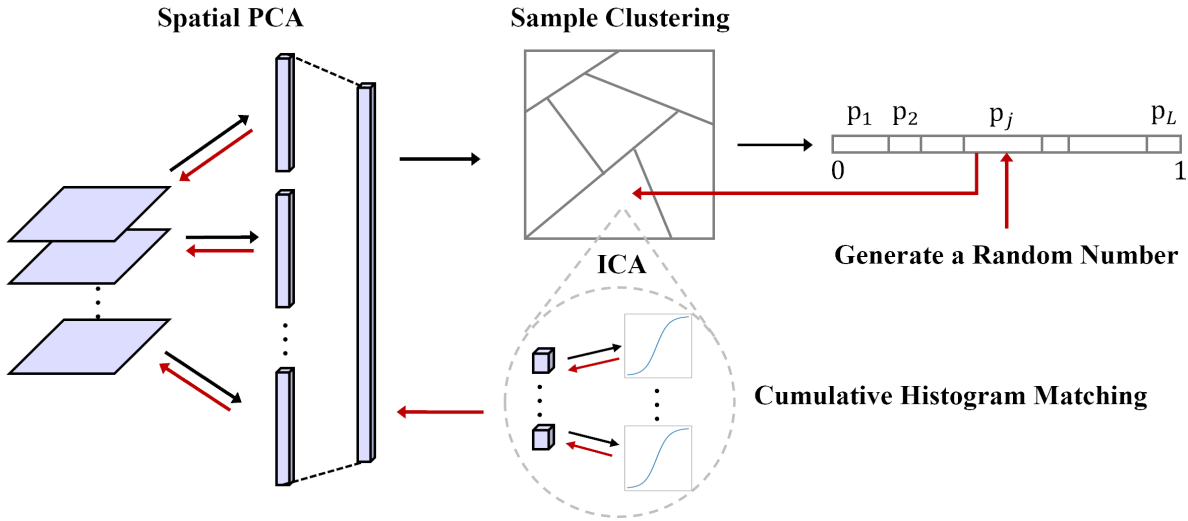}
\caption{Illustration of seed image generation in the lowest-dimensional subspace. }\label{fig:seed_analysis}
\end{figure}

In the training phase, we conduct the following four steps to learn
the sample distributions in the lowest dimension furthermore, as illustrated in Fig.~\ref{fig:seed_analysis}.
\begin{enumerate}
\itemsep 1ex
\item {\bf Spatial PCA.} There exist correlations between spatial
pixels $S_4$. They can be removed by applying PCA to the spatial
dimension of each channel, called spatial PCA. Components with
eigenvalues less than a threshold, $\gamma$, are discarded. The reduced
subspace and its dimension are denoted by $\tilde{S}_4$ and
$\tilde{D}_4$, respectively.  After a sequence of whitening operations,
elements of these vector samples are uncorrelated. However, they may
still be dependent. Furthermore, they are not Gaussian distributed. 
\item {\bf Sample Clustering.} Given training samples in
$\tilde{S}_4$, we perform k-means clustering on them to generate
multiple clusters so that the sample distribution of each cluster can be
simplified. This is especially essential for multi-modal sample
distributions. 

\item {\bf Independent Component Analysis (ICA).} We perform ICA 
in each cluster to ensure elements of vector samples are independent. 

\item {\bf Cumulative Histogram Matching.} We would like to match the
cumulative histogram of each independent component in a cluster with that
of a Gaussian random variable of zero mean and unit
variance~\cite{9306451,lei2021tghop}. 
\end{enumerate}

In the generation phase, we conduct the following steps, which are the
inverse of the operations as described above. 
\begin{enumerate}
\itemsep 1ex
\item {\bf Cluster Selection.} The probability of selecting a cluster is defined
by the ratio of the number of samples in that cluster and the total number of samples
in $\tilde{S}_4$. We randomly select a cluster based on its probability.
\item {\bf Sample Generation.} We generate a random variable using the
Gaussian density of zero mean and unit variance and map it to the
corresponding value of the sample distribution in the cluster via
inverse cumulative histogram matching. 
\item {\bf Inverse ICA.} It rebuilds dependency among elements of
random vectors.
\item {\bf Inverse Spatial PCA.} It rebuilds spatial correlations
among pixels of each channel.
\end{enumerate}

\vspace{1ex}
\subsection{Module 3: Low-to-High Dimension Expansion}\label{subsec:generation}

{\bf Recovering Discarded Details via LLE.} The generated sample in
$S_4$ contains only low-frequency (LF) components since high-frequency
(HF) components are discarded to simplify the seed generation procedure.
HF responses should be generated along the reverse direction to enhance
details. We adopt LLE~\cite{roweis2000nonlinear} to achieve this task.
LLE is a commonly used technique to build the correspondence between
low- and high-resolution images in image super-resolution
\cite{chang2004super} or restoration~\cite{huang2016visible}. Here, LLE
is used to adjust generated LF samples to ensure two things. First,
generated LF samples are located on the manifold of training LF samples.
Second, we determine the correspondence between samples of LF components
only and samples of both LF and HF components. LLE is implemented in
small regions of spatial resolutions 2x2 or 3x3. 

{\bf Neighborhood Coloring via inverse Saab Transform.} Finally, we
build the correlations among spatial pixels via the inverse Saab
transform, which can be interpreted as a coloring process.  The Saab
transform parameters are determined by PCA of AC components of a local
neighborhood. The parameters of the inverse transform can be derived
accordingly.

\section{Experiments} \label{sec:exp}

{\bf Experimental Setup.} We conduct experiments on three datasets:
MNIST, Fashion-MNIST and CelebA. They are often used for unconditional
image generation. MNIST and Fashion-MNIST contain gray-scale images
(i.e. $K_0=1$) while CelebA contains RGB color images.  To remove the
correlation between R, G, B three color channels, we perform pixel-wise
PCA to decouple them, yielding three uncorrelated channels denoted by P,
Q and R. We discard the R channel that has the smallest eigenvalue to
reduce the dimension. To recover the RGB channels, we apply LLE
conditioned on generated P and Q channels. As a result, $K_0=2$ for
CelebA. Hyper-parameters $(K_{1,l}, K_{1,h}, K_{2,l}, K_{2,h})$ are set
to (2, 1, 4, 3), (2, 2, 4, 4) and (3, 1, 4, 4) for MNIST, Fashion-MNIST and
CelebA, respectively. They are chosen to ensure the gradual dimension
transition between two successive subspaces.  The eigenvalue threshold,
$\gamma$, is set to 0.01, 0.01 and 0.03 for MNIST, Fashion-MNIST and
CelebA, respectively. The number of nearest neighbors in LLE is
adaptively chosen and upper bounded by 3. For CelebA, since the number
of training samples of LLE from $S_1$ to $S_0$ is high, we perform LLE
at one location at a time.

\begin{table}[tb]
\begin{center}
\caption{Comparison of FID scores of the GenHop model and representative
adversarial and non-adversarial models. The lowest FID scores are
shown in bold while the second lowest FID scores are underlined.}\label{table:fid}
\begin{tabular}{lccc} \hline
           &  MNIST &  Fashion & CelebA    \\ \hline
MM GAN~\cite{goodfellow2014generative}    &  9.8 & 29.6 & 65.6 \\ 
NS GAN~\cite{goodfellow2014generative}     &  6.8 & 26.5 & 55.0 \\ 
LSGAN~\cite{mao2017least}      &  7.8 & 30.7 & 53.9 \\
WGAN~\cite{arjovsky2017wasserstein}       &  \underline{6.7} & 21.5 & 41.3 \\
WGAN-GP~\cite{gulrajani2017improved}    &  20.3 & 24.5 & $\mathbf{30.0}$ \\
DRAGAN~\cite{kodali2017convergence}     &  7.6 & 27.7  & 42.3 \\
BEGAN~\cite{berthelot2017began}      &  13.1 & 22.9  &  38.9 \\  \hline
VAE~\cite{Kingma2014}        &  23.8 & 58.7 & 85.7 \\ 
GLO~\cite{bojanowski2018optimizing} &  49.6 & 57.7  & 52.4 \\ 
GLANN~\cite{hoshen2019non} &  8.6 & $\mathbf{13.0}$ & 46.3 \\  \hline
Ours (GenHop)     & $\mathbf{5.1}$ & \underline{18.1} & \underline{40.3} \\\hline
\end{tabular}
\end{center}
\end{table}

{\bf Performance Comparison.} We compare the performance of GenHop with
several representative DL-based generative models in Table
\ref{table:fid}. The performance metric is the Fréchet Inception
Distance (FID) score. It is commonly used since
both diversity and fidelity of generated images are considered.  By
following the procedure described in~\cite{lucic2018gans}, we extract
the embedding of 10K generated and 10K real images from the test set
obtained by the Inception network and fit them into two multivariate
Gaussians, respectively. The difference between the two Gaussians are
measured by the Fréchet distance with their mean vectors and covariance
matrices. A smaller FID score means better performance. The FID scores
of representative GAN-based models (listed in the first section of Table
\ref{table:fid}) are collected from~\cite{lucic2018gans} while those of
non-adversarial models (the second section) are taken
from~\cite{hoshen2019non}.  We see from the table that Genhop has the
best FID score for MNIST and the second best for Fashion-MNIST and
CelebA.  Our method outperforms almost all other DL-based benchmarking
methods.

\begin{figure*}[tb]
\centering
\includegraphics[width=0.93\linewidth]{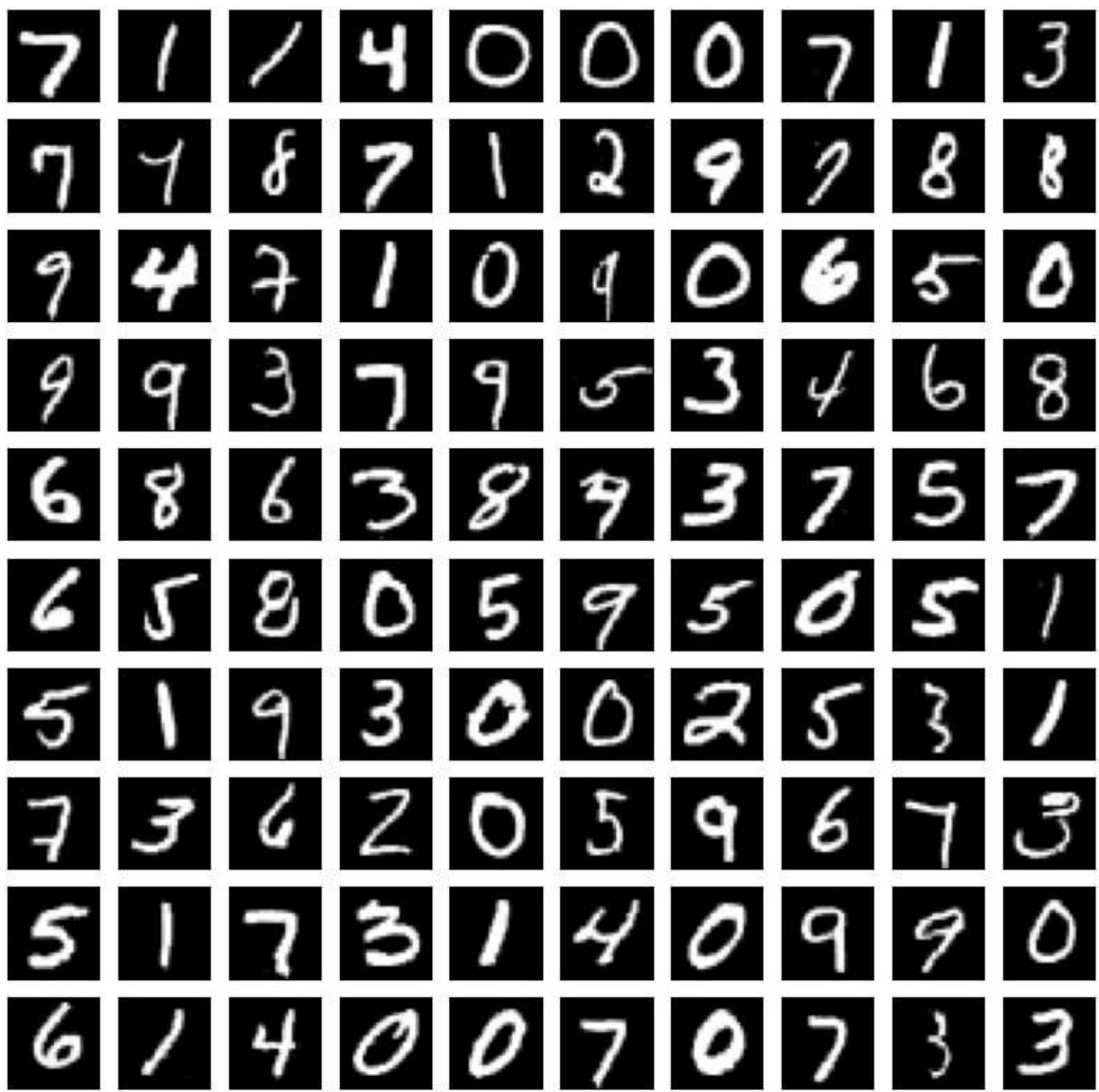} 
\caption{Exemplary images generated by GenHop with training samples 
from the MNIST dataset.} \label{fig:genimg_MNIST}
\end{figure*}

\begin{figure*}[tb]
\centering
\includegraphics[width=0.93\linewidth]{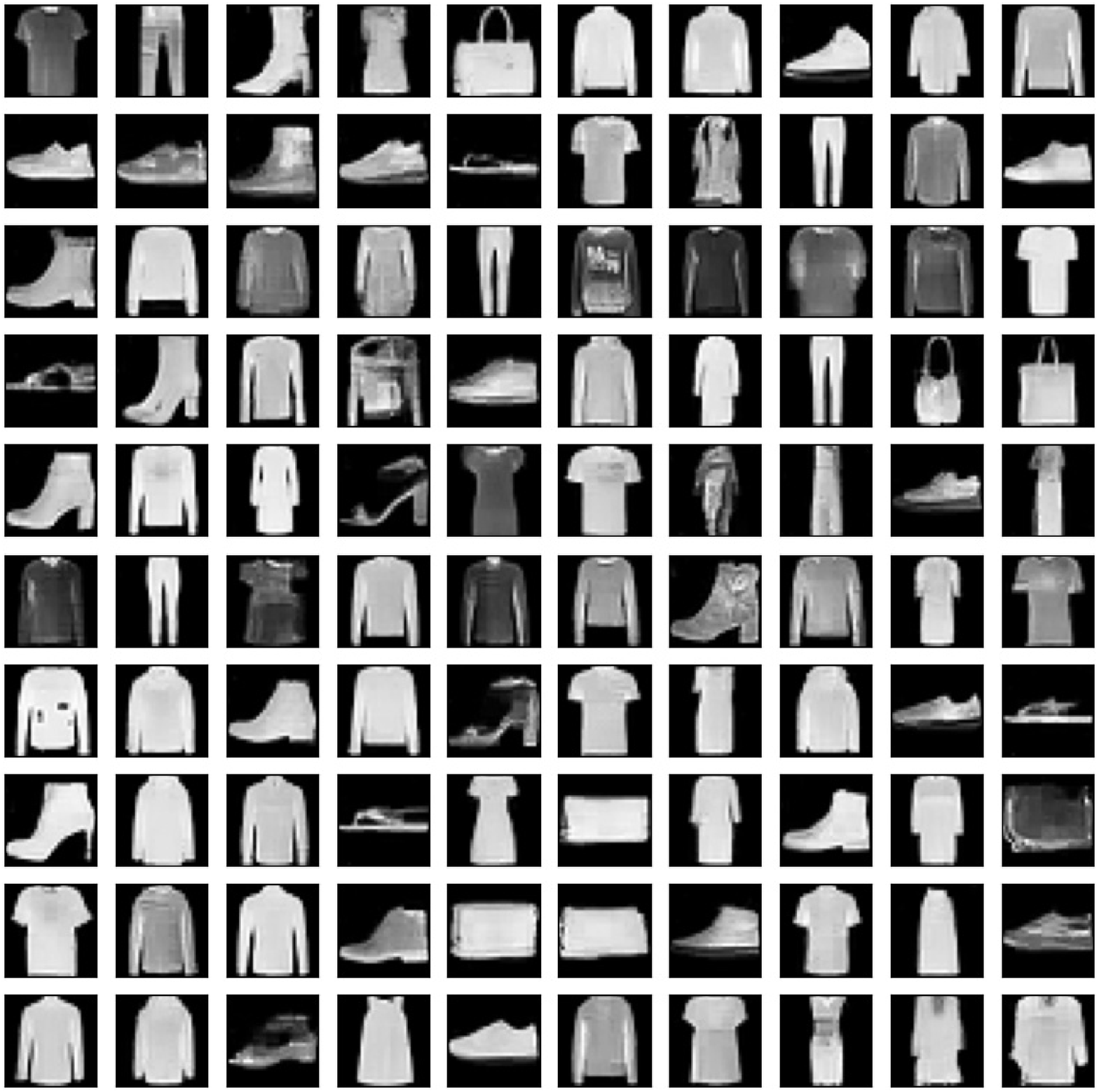}
\caption{Exemplary images generated by GenHop with training samples 
from the Fashion-MNIST dataset.} \label{fig:genimg_FMNIST}
\end{figure*}

\begin{figure*}[tb]
\centering
\includegraphics[width=0.93\linewidth]{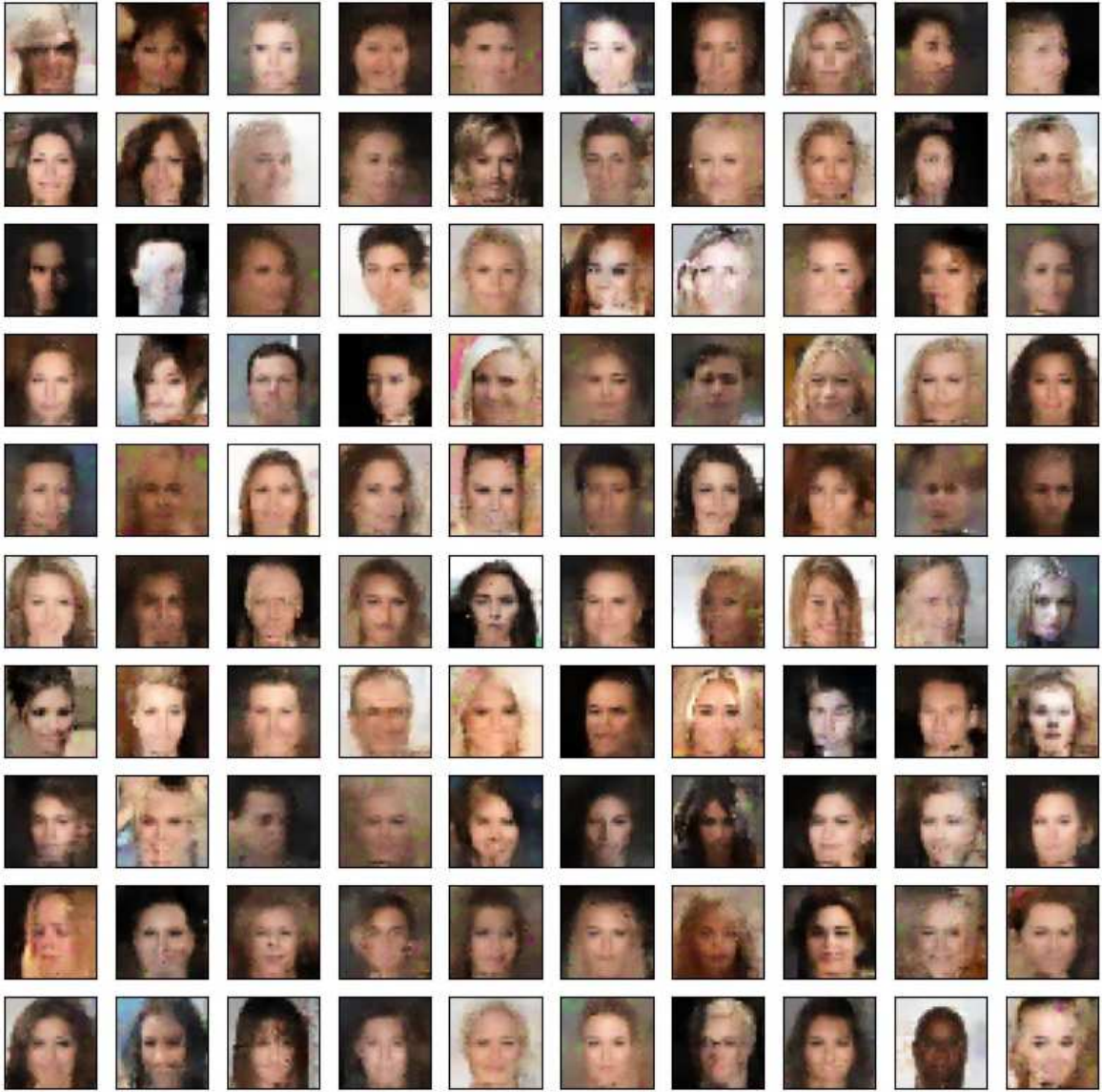} 
\caption{Exemplary images generated by GenHop with training samples 
from the CelebA dataset.} \label{fig:genimg_celebA}
\end{figure*}

{\bf Generated Exemplary Images.} Some exemplary images generated by
GenHop are shown in Fig.~\ref{fig:genimg_MNIST}, Fig.
\ref{fig:genimg_FMNIST} and Fig. \ref{fig:genimg_celebA} for visual
quality inspection.  For MNIST, the structure of digits is well captured
by GenHop with sufficient diversity. For Fashion-MNIST, GenHop generates
diverse examples for different classes. Fine details such as texture on
shoes and printing on T-shirts can be synthesized naturally. For CelebA,
the great majority of samples generated by GenHop are semantically
meaningful and realistic.  The color of some generated objects is not
natural. This failure is attributed to the lack of global information
since LLE is only performed in a small region of an image.

\section{Conclusion and Future Work}  \label{sec:conclusion}

A non-DL-based image generation method, called GenHop, was proposed in
this work. To summarize, GenHop conducted the following tasks: 1)
removing correlations among pixels in a local neighborhood via the Saab
transform, 2) discarding high frequency components for dimension
reduction, 3) generating seed images in the lowest dimensional space
using white Gaussian noise, 4) adding back discarded high frequency
components for dimension expansion based on LLE and 5) recovering
correlations of pixels via the inverse Saab transform. Tasks \#1 and \#2
can be done in multiple stages. Similarly, tasks \#4 and \#5 can also be
executed in multiple stages. GenHop achieved state-of-the-art
performance for MNIST, Fashion-MNIST and CelebA the datasets in FID
scores.  As future extension, it is desired to use GenHop to generate
images of higher resolution and more complicated content.  It is also
interesting to apply GenHop to the context of transfer learning (e.g.,
transfer between horses and zebras) and image inpainting.

\section*{Acknowledgment}

This research was supported by a gift grant from Mediatek. Computation
for the work was supported by the University of Southern California's
Center for High Performance Computing (hpc.usc.edu). 
\vfill\pagebreak

\bibliographystyle{unsrt}
\bibliography{references}

\begin{thebibliography}{10}

\bibitem{kuo2016understanding}
C-C~Jay Kuo.
\newblock Understanding convolutional neural networks with a mathematical
  model.
\newblock {\em Journal of Visual Communication and Image Representation},
  41:406--413, 2016.

\bibitem{kuo2017cnn}
C-C~Jay Kuo.
\newblock The cnn as a guided multilayer recos transform [lecture notes].
\newblock {\em IEEE signal processing magazine}, 34(3):81--89, 2017.

\bibitem{kuo2018data}
C-C~Jay Kuo and Yueru Chen.
\newblock On data-driven saak transform.
\newblock {\em Journal of Visual Communication and Image Representation},
  50:237--246, 2018.

\bibitem{kuo2019interpretable}
C-C~Jay Kuo, Min Zhang, Siyang Li, Jiali Duan, and Yueru Chen.
\newblock Interpretable convolutional neural networks via feedforward design.
\newblock {\em Journal of Visual Communication and Image Representation}, 2019.

\bibitem{chen2020pixelhop++}
Yueru Chen, Mozhdeh Rouhsedaghat, Suya You, Raghuveer Rao, and C-C~Jay Kuo.
\newblock Pixelhop++: A small successive-subspace-learning-based (ssl-based)
  model for image classification.
\newblock {\em arXiv preprint arXiv:2002.03141}, 2020.

\bibitem{goodfellow2014generative}
Ian Goodfellow, Jean Pouget-Abadie, Mehdi Mirza, Bing Xu, David Warde-Farley,
  Sherjil Ozair, Aaron Courville, and Yoshua Bengio.
\newblock Generative adversarial nets.
\newblock In {\em Advances in neural information processing systems}, pages
  2672--2680, 2014.

\bibitem{Kingma2014}
Diederik~P. Kingma and Max Welling.
\newblock {Auto-Encoding Variational Bayes}.
\newblock In {\em 2nd International Conference on Learning Representations,
  {ICLR} 2014, Banff, AB, Canada, April 14-16, 2014, Conference Track
  Proceedings}, 2014.

\bibitem{dinh2014nice}
Laurent Dinh, David Krueger, and Yoshua Bengio.
\newblock Nice: Non-linear independent components estimation.
\newblock {\em arXiv preprint arXiv:1410.8516}, 2014.

\bibitem{dinh2016density}
Laurent Dinh, Jascha Sohl-Dickstein, and Samy Bengio.
\newblock Density estimation using real nvp.
\newblock {\em arXiv preprint arXiv:1605.08803}, 2016.

\bibitem{hoshen2019non}
Yedid Hoshen, Ke~Li, and Jitendra Malik.
\newblock Non-adversarial image synthesis with generative latent nearest
  neighbors.
\newblock In {\em Proceedings of the IEEE Conference on Computer Vision and
  Pattern Recognition}, pages 5811--5819, 2019.

\bibitem{bojanowski2018optimizing}
Piotr Bojanowski, Armand Joulin, David Lopez-Pas, and Arthur Szlam.
\newblock Optimizing the latent space of generative networks.
\newblock In {\em International Conference on Machine Learning}, pages
  600--609, 2018.

\bibitem{li2018implicit}
Ke~Li and Jitendra Malik.
\newblock Implicit maximum likelihood estimation.
\newblock {\em arXiv preprint arXiv:1809.09087}, 2018.

\bibitem{chen2020pixelhop}
Yueru Chen and C-C~Jay Kuo.
\newblock Pixelhop: A successive subspace learning (ssl) method for object
  recognition.
\newblock {\em Journal of Visual Communication and Image Representation}, page
  102749, 2020.

\bibitem{chen2021defakehop}
Hong-Shuo Chen, Mozhdeh Rouhsedaghat, Hamza Ghani, Shuowen Hu, Suya You, and
  C-C~Jay Kuo.
\newblock Defakehop: A light-weight high-performance deepfake detector.
\newblock In {\em 2021 IEEE International Conference on Multimedia and Expo
  (ICME)}, pages 1--6. IEEE, 2021.

\bibitem{zhang2021anomalyhop}
Kaitai Zhang, Bin Wang, Wei Wang, Fahad Sohrab, Moncef Gabbouj, and C-C~Jay
  Kuo.
\newblock Anomalyhop: an ssl-based image anomaly localization method.
\newblock In {\em 2021 International Conference on Visual Communications and
  Image Processing (VCIP)}, pages 1--5. IEEE, 2021.

\bibitem{kadam2021r}
Pranav Kadam, Min Zhang, Shan Liu, and C-C~Jay Kuo.
\newblock R-pointhop: A green, accurate and unsupervised point cloud
  registration method.
\newblock {\em arXiv preprint arXiv:2103.08129}, 2021.

\bibitem{liu2021voxelhop}
Xiaofeng Liu, Fangxu Xing, Chao Yang, C-C~Jay Kuo, Suma Babu, Georges~El
  Fakhri, Thomas Jenkins, and Jonghye Woo.
\newblock Voxelhop: Successive subspace learning for als disease classification
  using structural mri.
\newblock {\em arXiv preprint arXiv:2101.05131}, 2021.

\bibitem{zhang2020pointhop++}
Min Zhang, Yifan Wang, Pranav Kadam, Shan Liu, and C-C~Jay Kuo.
\newblock Pointhop++: A lightweight learning model on point sets for 3d
  classification.
\newblock In {\em 2020 IEEE International Conference on Image Processing
  (ICIP)}, pages 3319--3323. IEEE, 2020.

\bibitem{zhang2020pointhop}
Min Zhang, Haoxuan You, Pranav Kadam, Shan Liu, and C-C~Jay Kuo.
\newblock Pointhop: An explainable machine learning method for point cloud
  classification.
\newblock {\em IEEE Transactions on Multimedia}, 2020.

\bibitem{zhang2020unsupervised}
Min Zhang, Pranav Kadam, Shan Liu, and C-C~Jay Kuo.
\newblock Unsupervised feedforward feature (uff) learning for point cloud
  classification and segmentation.
\newblock In {\em 2020 IEEE International Conference on Visual Communications
  and Image Processing (VCIP)}, pages 144--147. IEEE, 2020.

\bibitem{kadam2020unsupervised}
Pranav Kadam, Min Zhang, Shan Liu, and C-C~Jay Kuo.
\newblock Unsupervised point cloud registration via salient points analysis
  (spa).
\newblock In {\em 2020 IEEE International Conference on Visual Communications
  and Image Processing (VCIP)}, pages 5--8. IEEE, 2020.

\bibitem{manimaran2020visualization}
Abinaya Manimaran, Thiyagarajan Ramanathan, Suya You, and C-C~Jay Kuo.
\newblock Visualization, discriminability and applications of interpretable
  saak features.
\newblock {\em Journal of Visual Communication and Image Representation},
  66:102699, 2020.

\bibitem{tseng2020interpretable}
Tzu-Wei Tseng, Kai-Jiun Yang, C-C~Jay Kuo, and Shang-Ho Tsai.
\newblock An interpretable compression and classification system: Theory and
  applications.
\newblock {\em IEEE Access}, 8:143962--143974, 2020.

\bibitem{rouhsedaghat2020facehop}
Mozhdeh Rouhsedaghat, Yifan Wang, Xiou Ge, Shuowen Hu, Suya You, and C-C~Jay
  Kuo.
\newblock Facehop: A light-weight low-resolution face gender classification
  method.
\newblock {\em arXiv preprint arXiv:2007.09510}, 2020.

\bibitem{rouhsedaghat2021successive}
Mozhdeh Rouhsedaghat, Masoud Monajatipoor, Zohreh Azizi, and C-C~Jay Kuo.
\newblock Successive subspace learning: An overview.
\newblock {\em arXiv preprint arXiv:2103.00121}, 2021.

\bibitem{9306451}
Xuejing Lei, Ganning Zhao, and C.-C. Jay~Kuo.
\newblock Nites: A non-parametric interpretable texture synthesis method.
\newblock In {\em 2020 Asia-Pacific Signal and Information Processing
  Association Annual Summit and Conference (APSIPA ASC)}, pages 1698--1706,
  2020.

\bibitem{lei2021tghop}
Xuejing Lei, Ganning Zhao, Kaitai Zhang, and C-C~Jay Kuo.
\newblock Tghop: An explainable, efficient and lightweight method for texture
  generation.
\newblock {\em arXiv preprint arXiv:2107.04020}, 2021.

\bibitem{roweis2000nonlinear}
Sam~T Roweis and Lawrence~K Saul.
\newblock Nonlinear dimensionality reduction by locally linear embedding.
\newblock {\em science}, 290(5500):2323--2326, 2000.

\bibitem{chang2004super}
Hong Chang, Dit-Yan Yeung, and Yimin Xiong.
\newblock Super-resolution through neighbor embedding.
\newblock In {\em Proceedings of the 2004 IEEE Computer Society Conference on
  Computer Vision and Pattern Recognition, 2004. CVPR 2004.}, volume~1, pages
  I--I. IEEE, 2004.

\bibitem{huang2016visible}
Chun-Ting Huang, Zhengning Wang, and C.-C.~Jay Kuo.
\newblock Visible-light and near-infrared face recognition at a distance.
\newblock {\em Journal of Visual Communication and Image Representation},
  41:140--153, 2016.

\bibitem{mao2017least}
Xudong Mao, Qing Li, Haoran Xie, Raymond~YK Lau, Zhen Wang, and Stephen
  Paul~Smolley.
\newblock Least squares generative adversarial networks.
\newblock In {\em Proceedings of the IEEE international conference on computer
  vision}, pages 2794--2802, 2017.

\bibitem{arjovsky2017wasserstein}
Martin Arjovsky, Soumith Chintala, and L{\'e}on Bottou.
\newblock Wasserstein generative adversarial networks.
\newblock In {\em International Conference on Machine Learning}, pages
  214--223, 2017.

\bibitem{gulrajani2017improved}
Ishaan Gulrajani, Faruk Ahmed, Mart{\'\i}n Arjovsky, Vincent Dumoulin, and
  Aaron~C Courville.
\newblock Improved training of wasserstein gans.
\newblock In {\em NIPS}, 2017.

\bibitem{kodali2017convergence}
Naveen Kodali, Jacob Abernethy, James Hays, and Zsolt Kira.
\newblock On convergence and stability of gans.
\newblock {\em arXiv preprint arXiv:1705.07215}, 2017.

\bibitem{berthelot2017began}
David Berthelot, Thomas Schumm, and Luke Metz.
\newblock Began: Boundary equilibrium generative adversarial networks.
\newblock {\em arXiv preprint arXiv:1703.10717}, 2017.

\bibitem{lucic2018gans}
Mario Lucic, Karol Kurach, Marcin Michalski, Sylvain Gelly, and Olivier
  Bousquet.
\newblock Are gans created equal? a large-scale study.
\newblock In {\em Advances in neural information processing systems}, pages
  700--709, 2018.

\end{thebibliography}

\end{document}